\documentclass{aa}  

\usepackage{graphicx}
\usepackage{txfonts}
\usepackage{lipsum}
\usepackage{subcaption}         % necessary for continued figures, example in section 3
\usepackage{lscape}             % to rotate a single page table, example in appendix.
\usepackage{placeins}           % useful with \FloatBarrier, to keep 
\usepackage{hyperref}
\begin{document}

%%%%%%%%%%%%%%%%%%%%%%%%%%%%%%%%%%%%%%%%
% if you use custom commands in your title,
% ensure to check your title when submitting!
%%%%%%%%%%%%%%%%%%%%%%%%%%%%%%%%%%%%%%%%
   \title{Inferring the mass and size of 3I/ATLAS from its non-gravitational acceleration}
   %\subtitle{Subtitle}

%%%%%%%%%%%%%%%%%%%%%%%%%%%%%%%%%%%%%%%%
% Please separate each author with the \and command
%
% Please do not include ORCIDs next to author names.
% Only ORCIDs authenticated by individual authors in EDPS
% editorial system will be taken into account.
% ORCIDs included here will be removed.
%%%%%%%%%%%%%%%%%%%%%%%%%%%%%%%%%%%%%%%%

   \author{V. Thoss\inst{1,2,3}\fnmsep\thanks{Corresponding author: vthoss@mpe.mpg.de}
        \and A. Loeb\inst{4}
        \and A. Burkert\inst{1,2,3}
        }

   \institute{Universitäts-Sternwarte, Ludwig-Maximilians-Universität München, Scheinerstr. 1, 81679 Munich, Germany
   \and Excellence Cluster ORIGINS, Boltzmannstrasse 2, 85748 Garching, Germany
   \and Max-Planck Institute for Extraterrestrial Physics, Giessenbachstr. 1, 85748 Garching, Germany
   \and Astronomy Department, Harvard University, 60 Garden St., Cambridge, MA 02138, USA}

%\date{Submitted March 16, 2026}

% \abstract{}{}{}{}{}
% 5 {} token are mandatory
 
\abstract{Observations of the interstellar object 3I/ATLAS have revealed a strong production of gas and dust near perihelion, together with rapid brightening. The outgassing from the nucleus has led to a detectable non-gravitational acceleration. In this work, we combine models of the mass loss rate of water and carbon dioxide to derive the non-gravitational parameters and estimate the mass and size of 3I/ATLAS. In addition, we take into account a conservative constraint on the nucleus size from the active surface area required for sublimation. If the mass loss is dominated by the sublimation of CO$_2$, then the nucleus radius and mass are $R_{\rm 3I}=0.42\,\rm{km}$  and $M_{\rm 3I}=1.6\times10^{11}\,\rm{kg}$, assuming a density of $\rho=0.5\,\rm{g\,cm}^{-3}$ and an asymmetry factor of $\zeta=0.5$. This estimate is consistent with the lower bound from the active surface and independently supported by the slight preference of the orbital fit for a $a_{\rm ng}(r)\sim 1/r^2$ scaling of the non-gravitational acceleration. Models that cover the range of reported water production near perihelion give $R_{3I}=0.74-1.15\,\rm{km}$ and $M_{\rm 3I}=8.5-32\times10^{11}\,\rm{kg}$ but require a cometary surface that is in tension with the estimate from the rocket effect. Therefore, our results indicate that a large fraction of water sublimation is occurring in the coma and that CO$_2$ dominates sublimation on the surface. The nucleus radius that we obtain is much smaller than a recent photometric estimate of $R_{\rm 3I}\sim 1.3\,\rm{km}$, which could be resolved if CO$_2$ production is larger than observed or if the density of 3I/ATLAS is significantly lower than assumed. An overall lighter nucleus of 3I/ATLAS might be favored based on its recently claimed origin from a metal-poor environment and the corresponding mass budget of interstellar objects.}


   \maketitle
   \nolinenumbers

\section{Introduction}

On July 1, 2025, the Asteroid Terrestrial-impact Last Alert System (ATLAS) discovered the object C/2025 N1, later renamed to 3I/ATLAS. Follow-up and pre-discovery observations confirmed its hyperbolic orbit, with $e\approx 6.14$ and $r_p\approx 1.356\,\rm{AU}$ \citep{3AtlasJPL}, verifying its interstellar origin. Its velocity at infinity of $v_\infty\approx 58.0\,\rm{km\,s}^{-1}$ is large in comparison to the other two known interstellar objects, 1/Oumuamua ($v_\infty\approx26.4\,\rm{km\,s}^{-1}$, \cite{1IJPL}) and 2/Borisov ($v_\infty\approx32.3\,\rm{km\,s}^{-1}$, \cite{2IJPL}), although consistent with expectations from theoretical models \citep{oxford}. 

The coma of 3I/ATLAS dominates its overall brightness, making it difficult to directly determine the size of the nucleus. Using observations from the Hubble Space Telescope (HST), \cite{jewitt_hubble_2025} obtained an upper limit on the nucleus size of $R_{\rm 3I}<2.8\,\rm{km}$ by fitting the surface brightness profile. They also computed a lower limit of $R_{\rm 3I}>0.22\,\rm{km}$ based on the surface area required to supply the observed dust production rate through carbon monoxide sublimation. A complementary work by \cite{ScarmatoNucleus} found $R_{\rm 3I}=0.16-2.8\,\rm{km}$ using the HST data and applying bicubic resampling. Recently, \cite{Hui} have claimed a successful extraction of the nucleus with HST, reporting an effective radius of $R_{\rm 3I}=1.3\pm0.2\,\rm{km}$.

On its approach to perihelion, 3I/ATLAS has continuously brightened, displaying increasing cometary activity \citep{jewitt_preperihelion_2025}. Several studies have inferred the production rates of water, carbon dioxide, carbon monoxide, and other volatiles. The results suggest an initially carbon-dominated production at a heliocentric distance $r\gtrsim 2-3\,\rm{AU}$ \citep{cordiner_jwst_2025,lisse_spherex_2025}. Close to perihelion, a rapid brightening of 3I/ATLAS, together with large rates of water production have been reported \citep{zhang_rapid_2025,crovisier,Combi}. However, there is good evidence to suggest that a significant part of the water sublimation is occurring from icy grains in the coma instead of the nucleus \citep{cordiner_jwst_2025,yang_spectroscopic_2025,Li}. In this work, we summarize the observational results and use them to inform models of the overall mass loss rate $\dot{M}_{\rm 3I}$ of 3I/ATLAS. Due to momentum conservation, the mass loss from the nucleus causes a non-gravitational acceleration $a_{\rm ng}=\zeta\dot{M}v/M$. The magnitude of the acceleration depends on the outflow velocity $v$ of the gas or dust as well as the degree of collimation $\zeta$. In this work, we model $\dot{M}_{\rm 3I}$ based on the measured production rates and infer $a_{\rm ng}$ from the astrometric data published to the Minor Planet Center (MPC).

This paper is structured as follows: In Section~\ref{sec:production} we discuss the observational data on the gas and dust production of 3I/ATLAS and how they inform our empirical models of the mass loss rate. Section~\ref{sec:nongrav} explains how we compute the non-gravitational acceleration based on these models. The methods used to obtain the orbital fit of 3I/ATLAS are presented in Section~\ref{sec:orbitfit}, followed by an analysis of the sensitivity of the non-gravitational parameters to the weighting of the astrometric data in Section~\ref{sec:sensitivity}. In Section~\ref{sec:results} we present our estimates for the mass and size of 3I/ATLAS, discuss how they depend on our assumptions, and compare them to results from other work. We conclude with a summary in Section~\ref{sec:summary}.

\section{Gas and dust production of 3I/ATLAS}
\label{sec:production}

Numerous observations of 3I/ATLAS have been conducted at different stages of its trajectory through the solar system. Using spectroscopic measurements, production rates $Q$ of various molecules have been derived. In this section, we discuss the currently available data and how they guide our theoretical models of the overall mass loss rate of 3I/ATLAS. Throughout this paper, $r$ denotes the heliocentric distance of 3I/ATLAS, while $R$ denotes its physical radius.

\begin{figure*}
    \centering
    \includegraphics[width=0.95\linewidth]{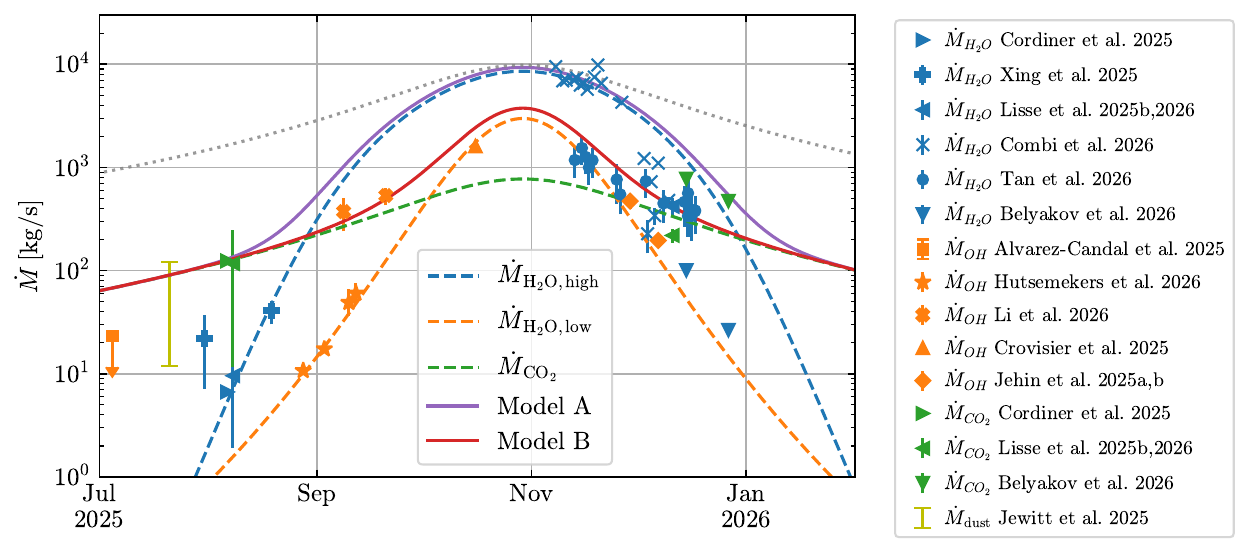}
    \caption{Compilation of observational data on the mass loss rates $\dot{M}$ of H$_2$O (blue), OH (orange) and CO$_2$ (green) for 3I/ATLAS. Individual references are given in the text. The colored dashed lines show empirical models for the mass loss rate, defined in Equations~\ref{eq:co2_model}, \ref{eq:Mdot_H2O_low}, and \ref{eq:h2o_model}. The solid pink and red curves display two models used in this work to describe the combined total emission rate from water and carbon dioxide, given by Equations~\ref{eq:mdotA} and \ref{eq:mdotB}. The gray dotted line illustrates the model implicitly assumed in the determination of the rocket effect by \cite{Hui}.}
    \label{fig:production}
\end{figure*}

In Figure~\ref{fig:production}, we compile all available data on the mass loss rates $\dot{M}=mQ$ of H$_2$O, CO$_2$, and OH, with $m$ being the respective mass of each molecule. The choice of these molecules is based on the assumption that the sublimation of H$_2$O and CO$_2$ dominate the total emission of 3I/ATLAS, supported by current data as we discuss below. Measurements of OH are included because they can serve as a proxy for the production rates of H$_2$O, with $Q_{\rm OH}\sim Q_{\rm H_2O}$.

Using the X-shooter instrument at the Very Large Telescope (VLT), \cite{alvarez-candal_x-shooter_2025} obtain upper limits on the production of OH at $r\approx 4.4\,\rm{AU}$ corresponding to a 3$\sigma$ limit of $\dot{M}_{OH}<23\,\rm{kg\,s}^{-1}$. At $r\approx 3.8\,\rm{AU}$, \cite{jewitt_hubble_2025} estimate an overall dust emission rate of $\dot{M}_{\rm dust}\sim1-120\,\rm{kg\,s}^{-1}$ based on HST observations. \cite{cordiner_jwst_2025} measure the production rate of CO$_2$ and H$_2$O at $r\approx 3.3\,\rm{AU}$ using the James Webb Space Telescope (JWST), finding $\dot{M}_{\rm CO_2}\approx 124\,\rm{kg\,s}^{-1}$ and an unusually high ratio $Q_{\rm CO_2}/Q_{\rm H_2O}\approx 7.6$ (corresponding to $\dot{M}_{\rm CO_2}/\dot{M}_{\rm H_2O}\approx 18.6$) compared to comets in the solar system at the same $r$, possibly due to cosmic ray processing \citep{maggiolo_interstellar_2025}. This result is in agreement with the study of \cite{lisse_spherex_2025} (updated in \cite{LisseUpdate}) that finds comparable rates of $Q_{\rm CO_2}$ for $r\approx 3.2\,\rm{AU}$ using SPHEREx. They derive an upper limit on the water production rate, which confirms the high ratio of $Q_{\rm CO_2}/Q_{\rm H_2O}$ at a similar heliocentric distance. Both studies also measure the production rate of CO, with \cite{lisse_spherex_2025} providing a 3-$\sigma$ upper limit on $Q_{\rm CO}$ that translates to $\dot{M}_{\rm CO_2}/\dot{M}_{\rm CO}>5.3$, while \cite{cordiner_jwst_2025} report the detection of CO production with $\dot{M}_{\rm CO_2}/\dot{M}_{\rm CO}\approx 7.2$. Other studies at later times in the orbit of 3I/ATLAS confirm the strong dominance of $Q_{\rm CO_2}$ over $Q_{\rm CO}$ \citep{Li}, and we therefore neglect the contribution of CO to the total mass loss rate.

The measurements of \citep{lisse_spherex_2025,cordiner_jwst_2025} together with the estimate provided by \cite{jewitt_hubble_2025} are consistent with an overall mass loss rate of $\dot{M}\sim100\,\rm{kg\,s}^{-1}$ at $r\sim3-4\,\rm{AU}$ and a carbon dioxide dominated coma. At this time, no further measurements of the CO or CO$_2$ production rate before perihelion have been published. \cite{Belyakov} and \cite{Lisse2} report production rates of CO$_2$ at $r\sim 2-3\,\rm{AU}$ post-perihelion, from JWST and SPHEREx observations, respectively. There is a discrepancy of around a factor of 3 between their values obtained in mid-December. Within these uncertainties, the values measured post-perihelion are roughly compatible with the CO$_2$ sublimation model of \cite{sekanina_sublimation_1992},
\begin{align}
    \dot{M}_{\rm CO_2}(r)&= 4.1\,\mathrm{kg\,s}^{-1}\left(\frac{r}{r_{\rm CO_2}}\right)^{-1.95} \times \nonumber\\
    &\times \exp\left(-1.73\left(\frac{r}{r_{\rm CO_2}}\right)^{1.5}\right)\left(1+\left(\frac{r}{r_{\rm CO_2}}\right)^{8.55}\right)^{-1.74}\,,
    \label{eq:co2_model}
\end{align}
with $r_{\rm CO_2}=20.2\,\rm{AU}$ from \cite{sekanina_sublimation_1992}. The normalization was obtained by a fit to the observational data and corresponds to a peak value of $\dot{M}_{\rm CO_2}\approx 770\,\rm{kg\,s}^{-1}$ at perihelion. The mass loss rate from Equation~\ref{eq:co2_model} scales approximately as $\dot{M}_{\rm CO_2}(r)\sim1/r^2$ for $r\lesssim 5\,\rm{AU}$ because the sublimation rate is limited by the energy from the solar radiation within the ice line. At larger distances, the predicted sublimation rate drops much steeper as a function of $r$ as radiative losses dominate the energy balance. The modeled production rate of CO$_2$ is shown as a green dashed line in Figure~\ref{fig:production}. \cite{jewitt_preperihelion_2025} find a power law scaling between the heliocentric magnitude $m_H$ of 3I/ATLAS and the heliocentric distance $r$ with a slope of $n=3.8\pm0.3$ for $r=1.8-4.6\,\rm{AU}$. As they note, this is consistent with a production rate that scales as $1/r^2$ for a dust-dominated coma. Closer to perihelion ($r=1.35-2\,\rm{AU}$), \cite{zhang_rapid_2025} and \cite{Eubanks} find a much steeper brightening of 3I/ATLAS, suggesting surging gas emission.

Further measurements of the production rates of H$_2$O and OH pre-perihelion have been conducted by \cite{xing_water_2025} (Swift Observatory), \cite{hutsemekers_extreme_2025} (VLT), \cite{crovisier} (Nançay Radio Observatory), and \cite{Li} (Purple Mountain Observatory). They are broadly consistent with a steep increase in water production towards perihelion, but there are significant discrepancies between the individual measurements. Observations with larger apertures report higher values of water production, demonstrated by \cite{Li}. This is indicative of an extended source of water production by icy grains in the coma that dominates the large-aperture observations. This interpretation is also supported by \cite{cordiner_jwst_2025}, who find an increase of water production as a function of the observed area around the nucleus (up to $R\sim 5000\,\rm{km}$). In addition, the spectroscopic analysis of \cite{yang_spectroscopic_2025} reveals absorption features consistent with a significant amount of water ice in the coma.

Post-perihelion measurements of the water production rates are carried out by \cite{Combi} and \cite{Tan} using the Solar and Heliosphere Observatory (SOHO), \cite{Lisse2} (SPHEREx) as well as \cite{Belyakov} (JWST). In addition, \cite{Jehin1} and \cite{Jehin2} measure $Q_{\rm OH}$ using TRAPPIST-North. As expected, the observations show a decreasing trend of the production rates. However, there is an unexplained discrepancy of almost an order of magnitude between the values obtained by \cite{Combi} and \cite{Tan} close to perihelion, both of which use SOHO observations. The discrepancy decreases for later measurements, when 3I/ATLAS was further away from the Sun. \cite{Tan} report a much shallower decrease in water production rate in comparison to \cite{Combi}. The water production rate obtained by \cite{Belyakov} on December 15 with the smaller aperture of JWST is lower by a factor of 3-4 than the values which \cite{Combi}, \cite{Lisse2}, and \cite{Tan} obtain at a similar time. This is consistent with an ongoing extended production of water in the coma after perihelion.

Sublimation of water from icy grains in the coma has no rocket effect on 3I/ATLAS. The previous emission of these grains from the surface of the comet is unlikely to dominate the rocket effect due to their small ejection velocities $v_{\rm grain}\ll v_{\rm gas}$ as they are dragged from the surface by the gas \citep{Xin}. The grains will only contribute significantly to the non-gravitational acceleration if $\dot{M}_{\rm grain}\gg \dot{M}_{\rm gas}$. This requires either grains with a very small ice fraction or very low levels of gas sublimation from the surface.\footnote{Note that in steady-state one has $\dot{M}_{\rm grain}=m_{\rm H_2O}Q_{\rm H_2O,coma}/f_{\rm ice}$, where $f_{\rm ice}$ is the mass fraction of the grains that is water ice.} Neither of these is supported by observations. Therefore, we should only account for the sublimation of water from the surface of the object. By analyzing the observational data and combining it with theoretical models, \cite{Li} find that the water production is initially strongly dominated by the sublimation in the coma, with $Q_{\rm H_2O,nuc}\ll Q_{\rm H_2O,coma}$ for $r\gtrsim 2\,\rm{AU}$, dropping to around equality, $Q_{\rm H_2O,nuc}\approx Q_{\rm H_2O,coma}$, at perihelion (although with very large uncertainties).

The large divergence between the different observational data makes it difficult to precisely assess the actual water production from 3I/ATLAS, which is further complicated by the uncertain fraction of sublimation in the coma. We therefore choose to provide two separate models that describe an upper and a lower limit of the water production rate according to the observations. For the lower bound, we describe $\dot{M}_{\rm H_2O}$ by a simple power law, given by
\begin{equation}
    \dot{M}_{\rm H_2O,low}(r)=3\times10^3\,\mathrm{kg\,s}^{-1}\left(\frac{r}{r_p}\right)^{-8.6}
    \label{eq:Mdot_H2O_low}
\end{equation}
which fits reasonably well to the data from \cite{hutsemekers_extreme_2025}, \cite{crovisier}, \cite{Jehin1}, \cite{Jehin2}, and \cite{Belyakov}. 

To describe the larger production rates measured by \cite{cordiner_jwst_2025}, \cite{lisse_spherex_2025}, \cite{xing_water_2025}, \cite{Li}, \cite{Combi}, and \cite{Lisse2}, we use a model based on \cite{marsden_comets_1973},
\begin{equation}
    \dot{M}_{\rm H_2O,high}(r)=2\times 10^3\,\mathrm{kg\,s}^{-1} \left(\frac{r}{r_{\rm H_2O}}\right)^{-2.15}\left(1+\left(\frac{r}{r_{\rm H_2O}}\right)^{5.09}\right)^{-4.61}\,,
    \label{eq:h2o_model}
\end{equation}
with $r_{\rm H_2O}=2.81\,\rm{AU}$. The model predicts $\dot{M}_{\rm H_2O}\sim10^4\,\rm{kg\,s}^{-1}$ at perihelion and $\dot{M}_{\rm H_2O}>\dot{M}_{\rm CO_2}$ for $r\lesssim2.7\,\rm{AU}$.

In both cases, the production rate of water exceeds that of carbon dioxide close to perihelion. This could explain the steep increase in magnitude observed for $r<2\,\rm{AU}$, first noted by \cite{zhang_rapid_2025}. Post-perihelion, the decrease of the magnitude of 3I/ATLAS with heliocentric distance has been shallower. This is shown by \cite{Tan}, who also claim that the water production rate decreased more slowly post-perihelion. However, due to the larger discrepancies between the production rates reported by different observers, we choose not to include an asymmetric model in this work. Furthermore, the water production rates obtained by observations with small apertures are reasonably well fitted by a symmetric power law (see Figure~\ref{fig:production}). On the other hand, the discrepancy between the reported production rates of carbon dioxide post-perihelion suggests the possibility of a delayed or asymmetric mass loss rate. In the future, with more available data, this could be a fruitful extension of our study. For now, we account for this uncertainty by having two different models for the water production rate, as we expect a more complex model to fall within these limits. \cite{Spada} studied asymmetric models for the non-gravitational acceleration and find a preference for a steeper power law slope post-perihelion vs. pre-perihelion. This is opposite to the results of \cite{Tan}, who find a shallower decrease after perihelion.

Production rates of other molecules have also been inferred, including HCN, CN, CH$_4$, CH$_3$OH, Fe, and Ni \citep{coulson_jcmt_2025,hutsemekers_extreme_2025,rahatgaonkar_vlt_2025,Roth,Lisse2,Belyakov,Paek}. However, their contributions to $\dot{M}$ are low, so we ignore them for our analysis. The emission of dust and cometary fragments could increase the mass loss rate of 3I/ATLAS beyond the gas emission captured in the previous discussion. However, the impulse imparted on the nucleus is likely subdominant to gas sublimation due to the much lower velocities. Therefore, we define our theoretical models based on the data of $\dot{M}_{\rm H_2O}$/$\dot{M}_{\rm OH}$ and $\dot{M}_{\rm CO_2}$, shown in Figure~\ref{fig:production}. 

Based on the previous discussion, we define two models to describe the dependence of $\dot{M}_{\rm 3I}$ on $r$. Model A, shown in Figure~\ref{fig:production} as a purple line, combines the model for $\dot{M}_{\rm CO_2}$ and $\dot{M}_{\rm H_2O, high}$, whereas model B, shown as a red line combines the rates of $\dot{M}_{\rm CO_2}$ and $\dot{M}_{\rm H_2O, low}$:
\begin{align}
    \label{eq:mdotA}
    \dot{M}_{\rm 3I,A}(r)&=\dot{M}_{\rm CO_2}(r)+\dot{M}_{\rm H_2O,high}(r)\\
    \dot{M}_{\rm 3I,B}(r)&=\dot{M}_{\rm CO_2}(r)+\dot{M}_{\rm H_2O,low}(r)\,.
    \label{eq:mdotB}
\end{align}

The integrated mass loss for these two models equates to $\Delta M_{\rm A}\approx5\times10^{10}\,\rm{kg}$ and $\Delta M_{\rm B}\approx1.6\times10^{10}\,\rm{kg}$. Note that this is an underestimate of the mass loss of 3I/ATLAS as it does not account for the mass loss in dust that may be of a similar order \citep{jewitt_hubble_2025,Gillan,Xin}.

\section{Non-gravitational acceleration}
\label{sec:nongrav}

Due to momentum conservation, the mass loss of 3I/ATLAS can induce a non-gravitational acceleration $a_{\rm ng}(r)$, defined by
\begin{equation}
    M_{\rm 3I}a_{\rm ng}(r)=\zeta\dot{M}_{\rm 3I}(r)v_{\rm gas}(r)\,,
    \label{eq:nongrav}
\end{equation}
where $v_{\rm gas}(r)$ is the (potentially $r$-dependent) velocity of the gas when it is emitted from the surface and $\zeta$ parametrizes the anisotropy of the outgassing. $\zeta=1$ corresponds to the maximum rocket effect, for a collimated outflow in a single direction and $\zeta=0$ to outgassing with zero net momentum transfer on the nucleus, e.g. for the case of exactly isotropic emission. Typically, values of $\zeta\sim0.5$ are assumed \citep{Rickman89,Sosa1}, but the precise value of $\zeta$ for 3I/ATLAS is unknown and potentially time-dependent. For the well-studied comet 67P/Churyumov-Gerasimenko, the non-gravitational acceleration fits the water production rate for $\zeta v_{\rm gas}\approx 480\,\rm{m\,s}^{-1}$ \citep{Kramer}\footnote{In this case, the mass of 67P is known independently, so $\zeta v_{\rm gas}$ remains as a free parameter.}. As this is comparable to the thermal velocity of H$_2$O for typical surface temperatures of $200\,\rm{K}$, it suggests $\zeta\sim0.5-1$ (\cite{Jewitt_67P} estimated $\zeta\sim0.5$ independently). It is beyond the scope of this work to estimate the value of $\zeta$ for 3I/ATLAS from the outgassing patterns (studied, e.g., in \cite{Roth} and \cite{Hoogendam}), so we will keep it as a free parameter, noting that $\zeta\sim0.1-1$ appears to be the plausible range. 

The gas velocity can be assumed to correspond roughly to the mean thermal velocity,
\begin{equation}
    \langle v_{\rm th}\rangle=\sqrt{\frac{8k_BT}{\pi m}}\,,
\end{equation}
at the surface of 3I/ATLAS. Energy-limited sublimation within the ice line implies $T\approx 200\,\rm{K}$ for H$_2$O and $T\approx120\,\rm{K}$ for CO$_2$ \citep{Skorov67P,Marschall67P,Bjorn67P}, corresponding to $v_{\rm H_2O}\approx 500\,\rm{m\,s}^{-1}$ and $v_{\rm CO_2}\approx 240\,\rm{m\,s}^{-1}$, respectively. At larger heliocentric distances, when the surface temperature is determined by radiative equilibrium, the velocities will be smaller. However, since $\dot{M}$ and hence the non-gravitational acceleration is much smaller further out from the Sun, we can safely neglect this for our analysis. Note that $v_{\rm th}\sim \sqrt{T}\sim r^{-1/4}$ for radiative equilibrium, but $\dot{M}$ typically decreases much more steeply with $r$.

By combining the stated values for $v_{\rm H_2O}$ and $v_{\rm CO_2}$ together with the models for the mass loss rate (Equations~\ref{eq:mdotA} and \ref{eq:mdotB}), we can formulate the radial dependence of the non-gravitational acceleration,
\begin{align}
    a_{\rm ng,A}(r)&=\gamma_{\rm A}(v_{\rm CO_2}\dot{M}_{\rm CO_2}(r)+v_{\rm H_2O}\dot{M}_{\rm H_2O,high}(r))\nonumber\\
    a_{\rm ng,B}(r)&=\gamma_{\rm B}(v_{\rm CO_2}\dot{M}_{\rm CO_2}(r)+v_{\rm H_2O}\dot{M}_{\rm H_2O,low}(r))\,.
    \label{eq:nongrav_models}
\end{align}
The constant of proportionality $\gamma_{\rm A/B}=\zeta/M_{\rm 3I}$ is the parameter that we infer in this work by confronting the model for $a_{\rm ng}(r)$ to astrometric data that constrain the orbit of 3I/ATLAS. In our analysis we will also consider a purely CO$_2$-driven sublimation following a $1/r^2$ scaling,
\begin{equation}
    a_{\rm ng,CO_2}=7.7\times 10^2\,\mathrm{kg\,s}^{-1}\gamma_{\rm CO_2}v_{\rm CO_2}(r/r_p)^{-2}\,.
    \label{eq:co2_ang}
\end{equation}
where the normalization is chosen to match Equation~\ref{eq:co2_model} near perihelion. Note that Equation~\ref{eq:co2_model} \citep{sekanina_sublimation_1992} has a slightly shallower slope (-1.95) and falls off rapidly outside the ice line. For our purposes, this difference has no relevant effect. To ensure maximum comparability with other studies, we choose a $a_{\rm ng}(r)\sim 1/r^2$ scaling for the case of CO$_2$-driven sublimation.

\section{Inference of orbital parameters}
\label{sec:orbitfit}

\begin{figure*}[t]
    \centering
    \includegraphics[width=0.95\linewidth]{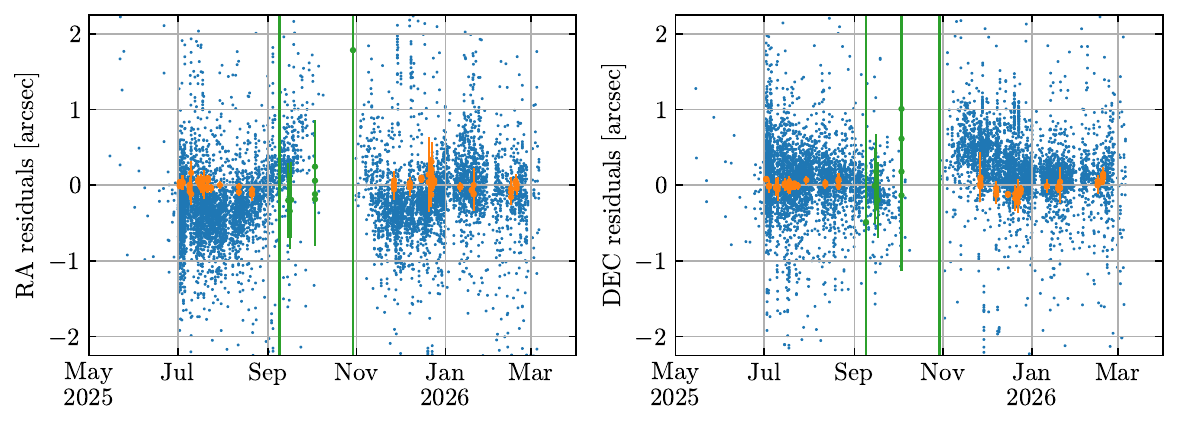}
    \caption{Residuals in RA and DEC between the data from the MPC and the orbital solution JPL \#54. The orange points indicate high-fidelity observations from large telescopes (specified in Section~\ref{sec:sensitivity}), including the reported residuals. Green color indicates measurements from interplanetary spacecraft (Trace Gas orbiter, Lucy spacecraft, Psyche spacecraft) that are valuable in constraining the orbit due to the triangulation.}
    \label{fig:mpc_jpl_res}
\end{figure*}

In order to estimate the mass of 3I/ATLAS, we infer its non-gravitational acceleration based on its orbital motion through the solar system. Using ASSIST \citep{assist}, an extension of the REBOUND code \citep{rebound} that provides ephemeris-quality test particle integration, we reconstruct the orbit of 3I/ATLAS given our model of its non-gravitational acceleration $a_{\rm ng}(r)$. ASSIST incorporates the position and velocities of the Sun, the planets, and 16 most massive asteroids based on JPL's DE441 ephemeris \citep{DE441}. In addition, ASSIST accounts for the most important finite-size effects and relativistic corrections, altogether providing excellent agreement with JPL's small-body orbit software, accessible through the Horizons web service.\footnote{https://ssd.jpl.nasa.gov/horizons} The orbit is integrated using IAS15, a 15th order Gauss-Radau integrator \citep{reboundias15}. As our models for $a_{\rm ng}(r)$ are different from the Marsden model \citep{marsden_comets_1973}, which is implemented in ASSIST, we modify the code to incorporate the radial dependence according to Equation~\ref{eq:nongrav_models}. 

The standard parameterization of the non-gravitational acceleration is defined by
\begin{equation}
    \vec{a}_{\rm ng}(r)=(A1\,\vec{\mathbf{r}}/r+A2\,\vec{\mathbf{t}}+A3\,\vec{\mathbf{h}}/h)g(r)\,,
\end{equation}
where $\vec{\mathbf{r}}$ is the vector pointing from the Sun to 3I/ATLAS, $\vec{\mathbf{h}}=\vec{\mathbf{r}}\times\vec{\mathbf{v}}$ gives the direction perpendicular to the orbital plane and $\vec{\mathbf{t}}=\vec{\mathbf{h}}\times\vec{\mathbf{r}}$ is perpendicular to $\vec{\mathbf{r}}$ but in the orbital plane. The radial dependence is incorporated into $g(r)$, which is defined by Equations~\ref{eq:nongrav_models} and \ref{eq:co2_ang}. Therefore, by inferring $A\equiv\sqrt{A1^2+A2^2+A3^3}$ we obtain $\gamma_{\rm A/B/ CO_2}$ and thus the mass of 3I/ATLAS. 

The orbital fit is obtained using the dataset collected by the MPC, containing 8086 observations from 2025-05-08 until 2026-03-06. In Figure~\ref{fig:mpc_jpl_res}, we display the residuals in RA and DEC between the data from the MPC and JPL's orbital solution \#54 (published Feb 19th). Notably, the highlighted data points that come from the largest telescopes with the best seeing and resolution agree well with the solution. However, the remaining observations are clearly not normally distributed around the solution, most notably in the residuals of the RA. If we trust the fewer data points with higher fidelity and thus the solution from JPL, this indicates significant systematic uncertainties at the level of 0.5 arcsec in many of the reported observations, possibly due to a tailward bias.

To understand the impact of these potential biases on the estimation of the mass and size of 3I/ATLAS, we choose several different methods of weighting the data, which are described in more detail in Section \ref{sec:sensitivity}. The most important result is that the magnitude $A$ of the non-gravitational acceleration is quite robustly estimated, changing only 5\% between the different weighting schemes considered. This is insignificant compared to the modeling uncertainty in the mass loss rate, the velocity of the gas, and the collimation factor $\zeta$. The values and confidence intervals quoted in Section~\ref{sec:results} are obtained by considering the range of values obtained by the different weighting schemes. However, we caution that this might still underestimate the actual uncertainty from the orbital fit.

\section{Sensitivity analysis of the orbital fitting procedure}

\label{sec:sensitivity}

Due to the systematic uncertainties present in the dataset from the MPC, it is necessary to carefully investigate the robustness of the inference of the non-gravitational parameters $A_i$ to the weighting and selection of the data. We use four different weighting schemes, which we describe in the following. In all cases, we account for over-observation bias by down-weighting data when an observatory reports more than four measurements in a single night. In this case, the residuals are scaled by a factor of $\sqrt{N/4}$, where $N$ is the number of observations in that night. In addition, we reject $>5\sigma$ outliers during the fit and iterate the fitting procedure until convergence is reached.

In the following, we list the four weighting schemes considered. When we refer to high-fidelity observations below, these include measurements from the following observatories: HST, VLT, Gemini North, Canada–France–Hawaii Telescope, Southern Astrophysical Research Telescope, and Apache Point Observatory. In total, 85 observations using these telescopes have been reported to the MPC.

\begin{enumerate}
    \item All data points are assumed to have the same residual $\sigma$ in RA and DEC. The value of $\sigma$ is given by an unbiased estimator, $\sigma=\sqrt{S/(n-p)}$, where $S$ is the sum of the square of the residuals, $n$ is the number of data points, and $p=9$ the number of orbital and non-gravitational parameters that are inferred. Note that $\chi_{\rm \nu}=1$ by construction in this case, so we do not report it. The values for $\sigma$ that we obtain are all very similar, $\sigma\approx0.63''$, regardless of the model.
    
    \item A residual of $\sigma=0.1''$ is assumed for the high-fidelity observations, unless a larger one is reported. For all other data, $\sigma=1''$ is assumed, unless a larger residual has been reported.
    
    \item Only observations which report a value for the seeing $s$, given in arcsec, are included. The residual in RA and DEC is computed as $\sigma=0.1''+0.1s$. If a larger uncertainty has been reported we use it instead. The relation has been chosen such that the residual for the lowest seeing values is at least $0.1''$ and the slope of $0.1$ has been found to lead to a well-constrained fit.
    
    \item Only high-fidelity observations are used for the fitting procedure and $\sigma=0.1''$ is assumed, unless a larger residual has been reported. We also include the data from the Trace Gas orbiter, Lucy spacecraft, and Psyche spacecraft due to their significance in constraining the orbit.
\end{enumerate}

\begin{table*}[h]
    \centering
    \begin{tabular}{c|ccc|c|c|c}
         Weighting scheme & $A_1$ [$10^{-8}$ AU d$^{-2}$] & $A_2$ [$10^{-8}$ AU d$^{-2}$] & $A_3$ [$10^{-8}$ AU d$^{-2}$] & $A$ [$10^{-8}$ AU d$^{-2}$] & $\chi_{\nu}$ & $d_{\rm JPL}$\\
         \hline
         1, Model A & $6.03\pm 0.05$ & $1.98 \pm 0.07$ & $-0.849 \pm 0.014$ & $6.40 \pm 0.05$ & – &–\\
         2, Model A & $5.84\pm 0.09$ & $2.00\pm 0.10$ & $-0.911\pm 0.021$ & $6.24\pm 0.07$ & 0.647 &–\\
         3, Model A & $6.06\pm0.09$ & $1.72\pm0.10$ & $-0.893\pm0.019$ & $6.36\pm0.07$ & 0.835 &–\\
         4, Model A & $6.06\pm0.10$ & $1.72\pm0.15$ & $-1.0\pm0.03$ & $6.39\pm0.08$ & 0.865 & –\\
         \hline
         1, Model B & $8.52\pm0.07$ & $2.42\pm 0.09$ & $-1.22\pm0.02$ & $8.94\pm 0.07$ & – &–\\
         2, Model B & $8.22\pm0.12$ & $2.61\pm 0.14$ & $-1.3\pm0.03$ & $8.73\pm 0.11$ & 0.649 & –\\
         3, Model B & $8.37\pm0.12$ & $2.23\pm 0.13$ & $-1.29\pm0.03$ & $8.76\pm0.10$ & 0.857 & –\\
         4, Model B & $8.39\pm0.14$ & $2.19\pm0.20$ & $-1.48\pm0.05$ & $8.80\pm0.11$ & 1.073 & –\\
         \hline
         1, CO$_2$ ($1/r^2$) & $4.48\pm0.05$ & $2.93\pm0.08$ & $-0.64\pm0.01$ & $5.39\pm0.04$ & – & 4.65\\
         2, CO$_2$ ($1/r^2$) & $4.4\pm 0.08$ & $2.64\pm 0.13$ & $-0.693\pm 0.016$ & $5.18\pm 0.06$ & 0.645 & 2.32\\
         3, CO$_2$ ($1/r^2$) & $4.73\pm0.08$ & $2.25\pm0.12$ & $-0.678\pm0.014$ & $5.28\pm0.06$ & 0.827 & 2.65\\
         4, CO$_2$ ($1/r^2$) & $4.80\pm0.10$ & $2.10\pm0.17$ & $-0.765\pm0.023$ & $5.29\pm0.06$ & 0.674 & 2.45\\
         \hline
         JPL \#51, ($1/r^2$) & $4.55\pm0.11$ & $1.77\pm0.14$ & $-0.620\pm0.022$ & $4.92\pm0.08$ & – & – \\ 
         JPL \#54, ($1/r^2$ + $DT$) & $5.32\pm0.12$ & $1.15\pm0.21$ & $-0.685\pm0.018$ & $5.49\pm0.09$ & – & – \\
         \cite{Spada}, ng1 & $3.78\pm 0.17$ & $3.86\pm 0.17$ & $-0.65\pm0.02$ & $5.44\pm0.17$ & – & –
    \end{tabular}
    \caption{Non-gravitational parameters $A_i$ obtained for the different models and weighting schemes. We also provide the goodness of fit parameters $\chi_\nu$ and, for the model with a $g(r)\sim1/r^2$ scaling, the difference $d_{\rm JPL}$ to the JPL solution, defined in the text. In the bottom rows, we quote the most recent solutions by JPL with and without a timing offset $DT$, as well as the results from \cite{Spada} for their model `ng1'.}
    \label{tab:sensitivity}
\end{table*}

In this order, the weighting schemes put progressively more weight on data points that have low seeing values and/or are from the selected telescopes listed above. In Table~\ref{tab:sensitivity}, we list the non-gravitational parameters $A_i$, the magnitude $A$, and the goodness of fit measured by the reduced chi-squared value $\chi_{\nu}^2=\frac{1}{n-p} \sum_i\frac{(m_i-y_i)^2}{\sigma_i^2}$. Here, $m_i$/$y_i$ denote the modeled/observed values of RA and DEC, $\sigma_i$ the residuals, $n$ the number of data points, and $p=9$ the number of fitted parameters. We also report $d_{\rm JPL}=\sqrt{(\vec{p}_{\rm CO_2}-\vec{p}_{\rm JPL})^T\mathbf{\Sigma}^{-1}(\vec{p}_{\rm CO_2}-\vec{p}_{\rm JPL})}$, which measures the statistical difference between the fit obtained for $g(r)\sim 1/r^2$ and the equivalent result from JPL, last reported in solution \#51 from Feb 13th (later solutions were obtained with an additional time offset $DT$ between perihelion and the maximum of $g(r)$). Here, $\vec{p}_{\rm CO_2}$ and $\vec{p}_{\rm JPL}$ denote the vector of the 9 parameters obtained by our analysis and JPL and $\mathbf{\Sigma}=\mathbf{\Sigma}_{\rm CO_2}+\mathbf{\Sigma}_{\rm JPL}$ is the sum of the respective covariance matrices. Note that the non-gravitational parameters are all obtained by normalizing $g(r_p)=1/r_p^2$ at perihelion, to ensure a meaningful comparison to the solutions from JPL and \cite{Spada}.

Table~\ref{tab:sensitivity} also explicitly lists the non-gravitational parameters for JPL solution \#51 from February 13, the latest one to assume a pure $g(r)\sim1/r^2$ scaling. We also report the most recent solution \#54 (from February 19), which includes the $DT$ parameter (given by $DT=9.5\pm1.4\,\rm{d}$). Finally, we also report the solution for the `ng1' model from \cite{Spada}, which also assumes a $g(r)\sim 1/r^2$ scaling without timing offset. We also want to note that the values of $A_i$ for our different models (A vs. B vs. CO$_2$) cannot be directly compared, given the different functional dependencies assumed for $g(r)$.

The non-gravitational parameters $A_i$ in Table~\ref{tab:sensitivity} vary at the level of around 5-20\% for the different weighting schemes, with the magnitude $A$ being very robust to treatment of the data. The strongest deviations are observed for the parameter $A_2$. This is in agreement with the recent analysis from \cite{Spada}, which finds that $A_2$ shows the largest sensitivity to data selection. While they find $A_1\approx A_2$ for a $g(r)\sim 1/r^2$ model, we obtain $A_1/A_2\approx 1.5-2$, depending on the weighting. On the other hand, JPL finds $A_1/A_2\approx 2.6$. Interestingly, our values of $A_2$ lie exactly between the results from JPL and \cite{Spada}.

From a statistical point of view, values of $A_i$ and $A$ for different weighting of the data are in slight tension. This is, however, not surprising and merely reflects that the statistical residuals from the fit do not fully capture the systematic uncertainty in the data. As expected, weighting schemes 2-4, which put more emphasis on the high-fidelity observations show better agreement with the solution from JPL. It should also be noted that all weighting schemes show the best goodness of fit for a $g(r)\sim1/r^2$ scaling and the worst for model B. However, this preference is very subtle – except for the case of weighting scheme 4, which contains much fewer data points and is thus more sensitive to changes in the model.

The tendency of the orbital data to prefer a more shallow dependence of the non-gravitational acceleration with heliocentric distance was also observed by \cite{Spada}, who found a very slight preference for $n=2$ when studying accelerations that follow $g(r)\sim 1/r^n$. We conducted a similar experiment, computing $\chi_{\nu}$ for the different weighting schemes as a function of the power law slope $n$. Consistently, $n\approx2$ is the preferred exponent, although the preference is subtle except for weighting scheme 4. For weighting scheme 2/3/4, $\chi_{\nu}$ increases by 1\%/7\%/76\%, going from $n=2$ to $n=8$ and by 0.3\%/2\%/30\%, going from $n=2$ to $n=4$. In essence, if we trust the much fewer but likely more accurate data points from the large telescopes, then a significant preference for $g(r)\sim 1/r^2$ emerges. This has interesting consequences, suggesting that the rocket effect might be dominated by the sublimation of CO$_2$ with negligible contribution by water production.

\section{Results and Discussion}
\label{sec:results}

In Table~\ref{tab:mass_estimates}, we provide the estimates for the mass of 3I/ATLAS for each model. In addition, we give the values of $A$ for a $g(r)$ that is normalized to $g(r_p)=1/r_p^2$ at perihelion, facilitating comparison to the values given by JPL. We also add, for comparison, the value of $A$ obtained by JPL solution \#51, which assumes a $g(r)\sim 1/r^2$-scaling. It is not surprising that $M_{\rm 3I,A}>M_{\rm 3I,B}>M_{\rm 3I,CO_2}$, as this simply reflects $\dot{M}_{\rm 3I,A}>\dot{M}_{\rm 3I,B}>\dot{M}_{\rm 3I,CO_2}$. Assuming a density of $\rho=0.5\,\rm{g\,cm}^{-3}$ and $\zeta=0.5$, we convert the mass estimates to effective radii of 3I/ATLAS. These are provided in Table~\ref{tab:radius_estimate}, which also lists the results from other studies. Note that the radius scales as $(\zeta /\rho)^{1/3}$ and is thus weakly dependent on the assumed density and outgassing asymmetry. In the following, we want to discuss these results in the context of previous studies.

\begin{table}[t]
    \centering
    \begin{tabular}{c|c|c}
        Model & $A$ [$10^{-8}$ AU d$^{-2}$] & $M_{\rm 3I}/\zeta $ [$10^{12}$ kg]\\ \hline
        A & $6.17-6.47$ & $6.29-6.59$ \\
        B & $8.62-9.01$ & $1.68-1.75$ \\
        CO$_2$ & $5.12-5.43$  & $0.31-0.33$ \\
        JPL \#51 & 4.92 $\pm$ 0.08 & –\\
    \end{tabular}
    \caption{Estimates of the non-gravitational acceleration and mass of 3I/ATLAS. $A$ is the non-gravitational acceleration, normalized at perihelion such that $g(r_p)=1/r_p^2$. The last row lists the value obtained by JPL solution \#51 (with equivalent assumptions as our CO$_2$ model) for comparison. The error ranges have been obtained by the sensitivity analysis described in Section~\ref{sec:sensitivity} in the Appendix.}
    \label{tab:mass_estimates}
\end{table}

\begin{table}[t]
    \centering
    \begin{tabular}{c|c|c}
        Source & Method & $R_{\rm 3I}$ [km] \\ \hline
        Model A (this work) & Non-grav. effect & 1.15\\
        Model B (this work) & Non-grav. effect & 0.74\\
        CO$_2$ (this work) &  Non-grav. effect & 0.42\\
        \cite{seligman_discovery_2025} & Magnitude & $<11\pm 1$\\
        \cite{jewitt_hubble_2025} & Nucleus extraction & $<2.8$\\
        \cite{chandler_nsf-doe_2025} & Surface brightness & $<6.3\pm0.8$\\
        \cite{ScarmatoNucleus} & Nucleus extraction & 0.16-2.8\\
        \cite{ForbesNG} & Non-grav. effect & 0.41-0.55\\
        \cite{EubanksNG} & Non-grav. effect & 0.28\\
       \cite{Hui} & Nucleus extraction & $1.3\pm 0.2$\\
        \cite{Hui} & Non-grav. effect & $1.5\pm 0.1$\\ 
    \end{tabular}
    \caption{Estimates of the size of 3I/ATLAS. The values obtained through the non-gravitational acceleration are computed assuming a density of $\rho=0.5\,\rm{g\,cm}^{-3}$ and $\zeta=0.5$. Values from other studies are rescaled to these values to facilitate comparisons. Photometric estimates are rescaled to a common value of the albedo of $p_V=0.04$. Note that the radius scales as $R\sim (\zeta/\rho)^{1/3}$ for non-gravitational estimates and $R\sim p_V^{-1/2}$ for photometric ones.}
    \label{tab:radius_estimate}
\end{table}

The difference in the values of the inferred masses for each model vastly exceeds the statistical uncertainties from the fitting procedure. This demonstrates that uncertainties in the modeling of the mass loss from the nucleus are currently dominating the errors in the estimation of the mass and size of 3I/ATLAS. The smallest mass and nucleus size is obtained if only CO$_2$-driven sublimation contributes to the rocket effect of 3I/ATLAS. In this case, we find $R_{\rm 3I}\approx0.42\,\rm{km}$. The difference to the value of $R_{\rm 3I}\approx0.55\,\rm{km}$ obtained by \cite{ForbesNG} for the same scenario is mostly a result of the difference in the assumed velocity of the sublimating carbon dioxide. Whereas they set $v=0.8\,\mathrm{km\,s}^{-1}(r/\rm{AU})^{-0.5}$ based on empirical relations of the gas velocity in the coma, we choose $v=v_{\rm th}=0.24\,\rm{km\,s}^{-1}$. This is lower because CO$_2$ sublimates at lower temperatures and is a heavier molecule than water. Also, the gas velocity is typically higher in the coma compared to when it left the nucleus \citep{Marschall67P}. The value of $R_{\rm 3I}\approx 0.28\,\rm{km}$ obtained by \cite{EubanksNG} is lower due to the larger inferred non-gravitational acceleration in their work, which is outdated and inconsistent with the analysis carried out by JPL, \cite{Spada}, and this work. Even though significant water production has been observed around perihelion, it is unclear what fraction of it is coming from the coma and thus not contributing to the rocket effect of 3I/ATLAS. Our finding, that a $g(r)\sim 1/r^2$ dependence is slightly preferred by the data (this was also noted by \cite{Spada}) could suggest that the contribution of water sublimation from the nucleus might be fully negligible compared to that of carbon dioxide, implying a nucleus size of $R_{\rm 3I}= 0.42\,\rm{km}$ under the stated assumptions.

If sublimation of water contributes significantly to the non-gravitational acceleration, then the nucleus size has to be larger, with $R_{\rm 3I}=0.74-1.15\,\rm{km}$. The range here encompasses the difference between the smallest and largest values of the water production that have been reported (see Figure~\ref{fig:production}). As discussed in Section~\ref{sec:production}, the largest values are likely somewhat or perhaps even strongly dominated by the sublimation from icy grains in the coma \citep{Li}. Even the smaller reported values may still contain a sizable contribution from the coma. Since this has no rocket effect on 3I/ATLAS and the ejection of those icy grains from the nucleus also contributes insignificantly to the torque (see Section~\ref{sec:production}), the lower value of $R_{\rm 3I}$ seems to be the more plausible estimate of the two. Both estimates are significantly smaller than the value of $R_{\rm 3I}=1.5\pm 0.1\,\rm{km}$ inferred by \cite{Hui} in their analysis of the non-gravitational acceleration. The main reason is that they take the peak value for the water production of $\dot{M}\sim 10^4\,\rm{kg\,s}^{-1}$ from \cite{Combi} together with $g(r)\sim 1/r^2$. The corresponding gray line in Figure~\ref{fig:production} shows that this strongly over-predicts the production of water before and after perihelion relative to observed values. In addition, the measurements of $Q_{\rm H_2O}$ at perihelion obtained by \cite{Combi} and \cite{Tan} differ by one order of magnitude, adding additional uncertainty at this time. Together with the evidence of a significant production of water in the coma, this estimate therefore seems unrealistic. As we discuss below, such a large water sublimation from the surface would also imply an active fraction of the cometary surface above unity and is thus ruled out.

\begin{figure*}[t]
    \centering
    \includegraphics[width=0.9\linewidth]{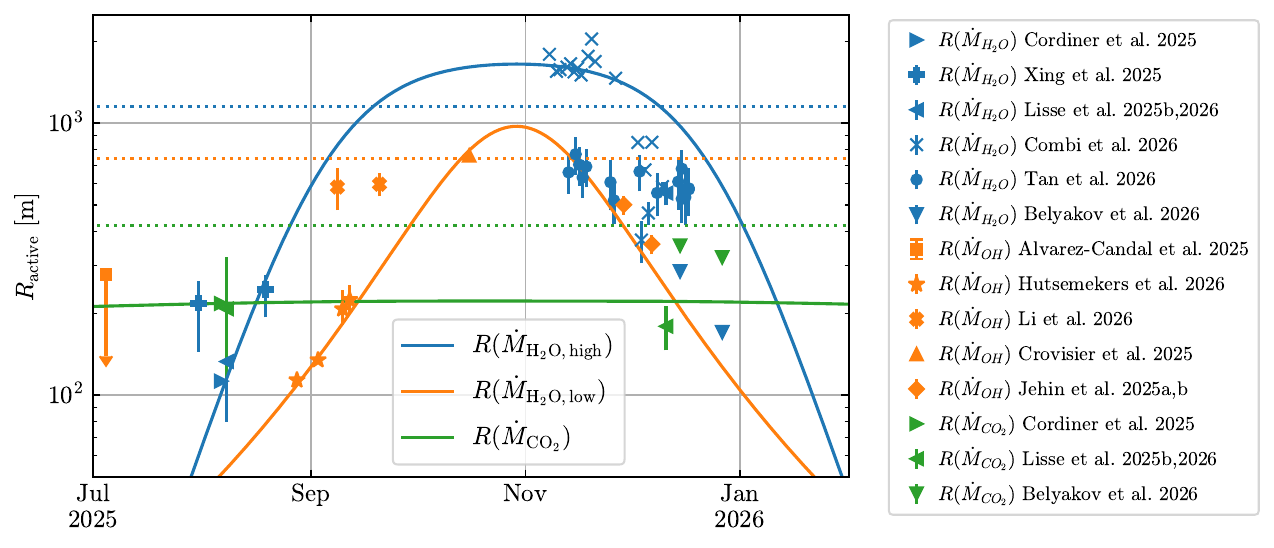}
    \caption{Minimum radius of 3I/ATLAS required to sustain the observed production rate with the entire surface being active. The solid lines correspond to the models described in Section~\ref{sec:production} and the dotted lines are the corresponding estimates of the radius of 3I/ATLAS from the non-gravitational effect.}
    \label{fig:radius_active}
\end{figure*}

Another way to gain insight into the size of 3I/ATLAS is to consider the surface area required to sustain the observed (and modeled) production rate of water and carbon dioxide through sublimation. Requiring that this surface does not exceed the actual surface of the object sets a relatively robust lower limit on the radius. Inside the ice line, where sublimation is energy-limited and radiative losses can be neglected, the flux from the solar radiation equals the latent heat flux from sublimation,
\begin{equation}
    \frac{S_\odot(1-A)}{r_\mathrm{AU}^2}\cos\theta=Lf\,,
\end{equation}
where $S_\odot=1360\,\rm{W\,m}^{-2}$ is the solar constant, $A=0.04$ the assumed albedo, $r_{\mathrm{AU}}$ the heliocentric distance in AU, $\theta$ the angle of the solar irradiation, $L$ the latent heat of the sublimating molecule and $f$ the mass loss through sublimation per time and surface area. The lower limit of the radius is then determined by requiring that $4\pi R^2f\geq\dot{M}$ and thus $R\geq \sqrt{\dot{M}/(4\pi f)}$. Using $L_{\rm H_2O}=2.84\times10^6\,\rm{J\,kg}^{-1}$ for water ice and $L_{\rm CO_2}=0.57\times10^6\,\rm{J\,kg}^{-1}$ for carbon dioxide, we obtain $f_{\rm H_2O}=4.6\times 10^{-4}\,\rm{kg}\,\rm{s}^{-1}\,\rm{m}^{-2}$ and $f_{\rm CO_2}=2.3\times 10^{-4}\,\rm{kg}\,\rm{s}^{-1}\,\rm{m}^{-2}$ at $r=1\,\rm{AU}$, respectively. To get the most conservative estimate of $R$, we assume $\theta=0$, i.e., subsolar temperature but isotropic emission. If hemispheric emission with the corresponding temperature is assumed, the lower bound for $R$ becomes a factor of two larger. In addition, comets typically have active fractions well below unity \citep{Hearn}. Note that since we assumed the energy-flux to be dominated by sublimation, this estimate is also overly conservative near and beyond the ice line, where the activity is expected to drop more steeply than $1/r^2$.

In Figure~\ref{fig:radius_active} we show the results from converting the observed values of $\dot{M}_{\rm H_2O}$ and $\dot{M}_{\rm CO_2}$ to lower bounds in the radius of 3I/ATLAS, denoted as $R_{\rm active}$. We also include the models for $\dot{M}_{\rm H_2O,low}$, $\dot{M}_{\rm H_2O,high}$ and $\dot{M}_{\rm CO_2}$ introduced in Section~\ref{sec:production}. In addition, the estimates of the effective size of the nucleus from the non-gravitational effect are shown in the corresponding colours as dotted lines. The sublimation of CO$_2$ requires a nucleus size of $R_{\rm active}\gtrsim 0.2\,\rm{km}$ that is lower than all the radii determined from the rocket effect. The scenario of dominant CO$_2$ sublimation at the observed rates, with negligible contribution from water production, is therefore consistent from this point of view. This is not the case for the other scenarios involving a significant contribution of water sublimation from the surface, described by model A and B. Both models require $R_{\rm active}$ around perihelion that is larger than the size obtained from the non-gravitational acceleration. The tension is worse for model A, which requires $R_{\rm active}\gtrsim1.5$ over an extended period of time (compared to $R_{\rm ng}=1.15\,\rm{km}$), whereas model B only exceeds the radius from the rocket effect very near the perihelion. In the latter case, the tension might be relaxed by a shallower increase near perihelion, between the observations of \cite{crovisier} and \cite{Tan}. However, we want to emphasize that the bound from the active surface is a very conservative estimate and that the true tension is likely stronger in both cases.

There are two possible ways to resolve this tension. On the one hand, the assumptions that were made in the computation of the radius from the non-gravitational acceleration may be inadequate. These include the assumed values for the gas velocity $v$, the asymmetry factor $\zeta$, and the comet density $\rho$. On the other hand, even the more conservative model B of water production and the corresponding observations may contain a significant fraction of sublimation from the coma. In this case, water sublimation from the surface might be similar or even negligible compared to CO$_2$. We can quantify how much we need to change our assumptions about $\rho$, $v$, and $\zeta$ to resolve the tension, by requiring $R_{\rm ng}\geq R_{\rm active}$. Since the non-gravitational estimates depend on the quantity $v\zeta/\rho$, let us denote $x=(v/v_{\rm th})(\zeta/0.5)(\rho/0.5\,\rm{g\,cm}^{-3})^{-1}$, with $x=1$ therefore describing our standard assumptions. From $R_{\rm active}=\sqrt{\dot{M}/(4\pi f)}$ and $R_{\rm ng}=(3\dot{M}v\zeta/4\pi\rho a_{\rm ng})^{1/3}$ we get $x_{\rm A}\geq 3$ and $x_{\rm B}\geq 2.5$. These are conservative bounds that are sensitive to the estimate $R_{\rm active}$. If we assume hemispheric emission or a more realistic active fraction of 25\%, $R_{\rm active}$ doubles and the bound on $x$ becomes eight times higher, giving $x_{\rm A}\geq 24$ and $x_{\rm B}\geq 20$. Models with significant water sublimation are therefore only allowed under the quite strong assumptions, extremely efficient sublimation over a large fraction of the surface with high gas velocities and/or a very low comet density, and are therefore disfavored. On the other hand, we get $x_{\rm CO_2}\geq 0.14$ and $x_{\rm CO_2}\geq 1.1$ for the conservative and more realistic estimate, demonstrating the consistency of the scenario of pure CO$_2$ sublimation. This is an important result and fits well to the previous observations that a $1/r^2$ model of the sublimation is (slightly) preferred by the fit. These results suggest that a large fraction of water sublimation occurs in the coma rather than from the nucleus. Note that throughout our analysis, we have assumed a spherical shape of 3I/Atlas. A highly nonspherical nucleus would allow for a larger surface area per volume and could soften the tension between $R_{\rm active}$ and $R_{\rm ng}$.

The previous considerations imply that the physically most plausible model for the mass loss rate is that of pure CO$_2$ sublimation, for which we obtain $R_{\rm 3I}=0.42\,\rm{km}$. This estimate is however much lower than the photometric estimate of $R_{\rm 3I}=1.3\pm0.2\,\rm{km}$ from \cite{Hui}. Raising our estimate to the lower bound of $R_{\rm 3I}=1.1\,\rm{km}$ would require $\dot{M}v\zeta/\rho$ to increase by a factor of 18. One explanation could be an underestimate of the CO$_2$ sublimation by observations in combination with a somewhat faster outgassing velocity or slightly lower cometary density. The constraint on CO$_2$ sublimation from the active fraction is less strong than that for water, leaving some room for a larger sublimation rate, at least under conservative assumptions. Another explanation is that the photometric estimate by \cite{Hui} is unreliable due to strong extinction around the nucleus. A lower size and mass of 3I/Atlas would also reduce the required mass budget of interstellar object, softening the tension with its claimed origin from a metal-poor environment \citep{CordinerIsotope,Salazar,Opitom,LoebIsotope}

Finally, we would like to point out that the scenario of a light nucleus with its outgassing dominated by CO$_2$ sublimation could imply a substantial mass loss during the encounter with the Solar System. Taking the more conservative model B as a lower bound and $\zeta=0.5$, 3I/ATLAS would have lost at least $5\%/\zeta$ of its initial mass. This is likely an underestimate given the additional mass loss in dust that we have not accounted for.

\section{Summary}
\label{sec:summary}

In this work, we have reviewed the observational data on the production rates of water and carbon dioxide of 3I/ATLAS to derive empirical models for the mass loss rate of the interstellar object. We consider three parameterizations: a purely CO$_2$-driven sublimation with a $1/r^2$-dependence and two models accounting for the contribution from water sublimation. These two models were fitted to the highest (A) and lowest (B) reported production rates, encompassing the range of uncertainty that is likely driven by an extended production in the coma. By combining these models with the astrometric data of 3I/ATLAS, we have derived the non-gravitational parameters and as a result, estimated the mass and size of the object. We have analyzed the sensitivity of our fitting procedure to different weighting and selection of the data, to understand the systematic uncertainties of the orbital solution. Finally, we have compared our size estimates of the nucleus to previous work and discussed additional limits from the active surface available for sublimation. Our results are summarized in the following.

\begin{itemize}
    \item Most of the available astrometric data shows biases and systematic uncertainties at the level of 0.5'' when compared to observations from the largest telescopes with the highest resolution.
    
    \item Despite these potential systematic uncertainties, the magnitude of the non-gravitational acceleration $A$ can be estimated quite robustly. Our different weighting schemes give results that differ by only 5\%. The strongest dependency on the treatment of the data was observed for $A_2$, in agreement with previous findings of \cite{Spada}.
    
    \item For all our models and weighting schemes, we find $A_1>A_2>A_3$. Our values for $A_1$ and $A_2$ lie between those obtained by \cite{Spada} and JPL. There is a subtle statistical preference towards the CO$_2$ model with a $1/r^2$ scaling, which becomes more pronounced when we only include the data from large telescopes and interplanetary spacecraft. When testing a generic power law scaling $g(r)\sim 1/r^n$ for the non-gravitational acceleration, we find that $n\approx 2$ is preferred compared to steeper slopes.
    
    \item The estimate for the mass of 3I/ATLAS is $M_{\rm 3I}/\zeta=0.31-0.33\times10^{12}\,\rm{kg}$ for a CO$_2$-only model, where $\zeta$ is the outgassing asymmetry factor. Including the contribution from water sublimation, we obtain $M_{\rm 3I}/\zeta=1.68-1.75\times10^{12}\,\rm{kg}$ and $M_{\rm 3I}/\zeta=6.29-6.59\times10^{12}\,\rm{kg}$ for the low and high limit of water sublimation from the nucleus. The difference between these values compared to the credible ranges from the fit shows that the errors in the estimates for the mass and size of 3I/ATLAS are currently strongly dominated by the uncertainty of the mass loss rate, rather than the astrometry.
    
    \item Assuming a density of $\rho=0.5\,\rm{g\,cm}^{-3}$ and $\zeta=0.5$, we estimate the nucleus radius of 3I/ATLAS to be $R_{\rm 3I}=0.42\,\rm{km}$ for the CO$_2$-driven sublimation, and $R_{\rm 3I}=0.74\,\rm{km}$ and $R_{\rm 3I}=1.15\,\rm{km}$ for the low (model B) and high (model A) limit of water sublimation. The value of $R_{\rm 3I}=0.42\,\rm{km}$ is consistent with a similar analysis by \cite{ForbesNG}, when we account for the different outgassing velocity assumed. We argue that the estimate of $R_{\rm 3I}=1.5\pm0.1\,\rm{km}$ from \cite{Hui} is too high because it assumes an unrealistically high mass loss rate, which exceeds almost all observed data points. In addition, there is good evidence that a large fraction of the observed water production is occurring in the coma and therefore without any rocket effect on the nucleus.

    \item We derive an additional constraint on the size of 3I/ATLAS by considering the surface area required to sustain the sublimation. Under the most conservative assumptions, this leads to a tension for the two models which include significant water production. The high sublimation rate would require a nucleus size that is too large to be compatible with the non-gravitational effect, unless we adopt extreme assumptions. A model with an even lower water production rate than our conservative model B might still be viable. This also implies that the sublimation of water from the surface of 3I/ATLAS likely does not significantly exceed that from CO$_2$. The surface area required for the sublimation of carbon dioxide is consistent with the corresponding nucleus size from the rocket effect.

    \item The combined evidence of a large fraction of water production occurring in the coma, the preference of the orbital solution for a purely CO$_2$-driven sublimation, and the constraints from the active fraction suggest that the rocket effect of 3I/ATLAS might be dominated by CO$_2$ sublimation throughout the orbit, with negligible contribution from water production. In this case the nucleus has an effective radius of $R_{\rm 3I}=0.42\,\rm{km}$, which calls the reliability of the photometric estimate of $R_{\rm 3I}=1.3\pm0.2\,\rm{km}$ into question. The estimates could be reconciled if the observations have underestimated the production rates of CO$_2$, perhaps in combination with a lower-than-usual comet density.
    
\end{itemize}

Our analysis has studied a range of possible scenarios for the outgassing from the surface of 3I/ATLAS and computed the corresponding mass and size estimates. The availability of more data on the production rates of water and carbon dioxide should help to narrow down the range of possible scenarios and thus improve estimates of the size and mass of 3I/ATLAS. Knowledge of the latter will be useful for our understanding of the origin and density of interstellar objects.

\begin{acknowledgements}
We thank Davide Farnocchia for providing us with the data for the orbital solution as well as for helpful comments and discussions. This research was supported in part by the Excellence Cluster ORIGINS which is funded by the Deutsche Forschungsgemeinschaft (DFG, German Research Foundation) under Germany’s Excellence Strategy - EXC-2094 - 390783311 (for V.T. and A.B.). This work was supported in part by the Black Hole Initiative and the Galileo Project at Harvard University (for A.L.).
\end{acknowledgements}

\bibliography{literature.bib}
\bibliographystyle{bibtex/aa.bst}

\end{document}